\documentclass[aps,prb,twocolumn,groupedaddress,showpacs,dvipdfmx]{revtex4-1}
\usepackage{amsmath}
\usepackage{bm}
\usepackage{braket}
\usepackage{amssymb}
\usepackage{graphicx}	
\usepackage{wrapfig}
\usepackage{setspace}
\usepackage{threeparttable}
\usepackage{longtable}
\usepackage{graphicx,color}
\begin{document}

\setstretch{1.0}

%\preprint{}

\title{First-principles evaluation of intrinsic, side-jump, and skew-scattering parts of anomalous Hall conductivities in disordered alloys}


\author{K. Hyodo and A. Sakuma}
%\email[]{hyodo@sakuma.apph.tohoku.ac.jp}
%\thanks{}
\affiliation{Department of Applied Physics, Tohoku University, Aoba 6-6-05, Aoba-ku, Sendai 980-8579, Japan}
\author{Y. Kota}
\affiliation{National Institute of Technology, Fukushima College, Iwaki, Fukushima 970-8034, Japan}

%\date{\today}

\begin{abstract}
We develop a first-principles procedure for the individual evaluation of the intrinsic, side-jump, and skew-scattering contributions to the anomalous Hall conductivity $\sigma_{xy}$.
This method is based on the different microscopic conductive processes of each origin of $\sigma_{xy}$ in the Kubo--Bastin formula.
We also present an approach for implementing this scheme in the tight-binding linear muffin-tin orbital (TB-LMTO) method with the coherent potential approximation (CPA).
The validity of this calculation method is demonstrated for disordered FePt and FePd alloys.
We find that the estimated value of each origin of $\sigma_{xy}$ exhibits reasonable dependencies on the electron scattering in these disordered alloys.
%al property of electron scattering from disordered alloys in calculated each origin corresponds with that predicted by the previous studies with respect to the electron scattering from the disorder alloys .
\end{abstract}

\pacs{72.10.Bg, 72.15.Gd, 75.47.Np}
%\keywords{}

\maketitle

\section{Introduction}
The spin-transport driven by the spin--orbit-interaction (SOI) is one of the new attractive fields of spintronics, which is called ``spin-orbitronics\cite{Kuschel2015}.''
The anomalous Hall effect (AHE)\cite{Hall1879,RevModPhys.82.1539} is a well-known SOI-driven transport phenomenon.
This effect results in a spin-dependent transverse electric current perpendicular to the external electric field and the magnetization of ferromagnets.
Recent interest in this phenomenon is related to its close connection with the spin Hall effect\cite{RevModPhys.87.1213}; this effect is expected to be applied in magnetic devices based on spin-orbitronics.

One of the perspectives of the AHE in metals is the multiple mechanisms of this effect. 
The intrinsic origin\cite{PhysRev.95.1154} provides a finite conductivity even in perfect crystals and originates from the effective magnetic field from the Berry phase of electrons\cite{doi:10.1143/JPSJ.71.19}.
On the other hand, the extrinsic origin needs electron scattering by impurities.
This origin is further categorized into the ``skew-scattering mechanism\cite{Smit1955877},'' which diverges in the clean limit, and the ``side-jump mechanism\cite{PhysRevB.2.4559},'' which has a finite value in the same limit.
The side-jump mechanism was recently reconsidered with regard to its connection with the Berry phase of electrons similar to the intrinsic origin\cite{PhysRevB.72.045346,PhysRevB.73.075318}.
Owing to the different physical properties for these mechanisms mentioned above, evaluation of the contributions of each mechanism individually is a productive approach for greater understanding of the AHE in ferromagnetic metals. %, which have the different physical properties mentioned above, have still seemed to cause confusion in the understanding of the AHE in metallic systems.
%It has been difficult to clarify the predominant origin of the AHE in metals.

In experimental approaches, the measured anomalous Hall conductivity $(\sigma_{xy})$ is often categorized into two parts using the distinct dependence of these origins on the longitudinal conductivity ($\sigma_{xx}$) as follows\cite{RevModPhys.82.1539}:
\begin{align}
\sigma_{xy} \simeq a\sigma_{xx} + b,\label{eq:00}
\end{align}
where $a\sigma_{xx}$ and $b$ are the contributions from the skew-scattering mechanism and the sum of the intrinsic and side-jump origins, respectively.
In dilute alloys with a large $\sigma_{xx}$, the first term in Eq. (\ref{eq:00}) is dominant, and $\sigma_{xy}$ almost behaves as being proportional to $\sigma_{xx}$.
However, the relative relationship between the contributions of the intrinsic and side-jump origins is unresolvable with this method.
First-principles studies have the potential to evaluate these origins in real systems separately from a microscopic viewpoint.
Actually, the contribution of the intrinsic origin was calculated in perfect crystals such as pure metals\cite{PhysRevLett.92.037204,PhysRevB.76.195109} and ordered alloys\cite{PhysRevB.67.174406,Fang92,PhysRevB.85.012405}.

Some recent theoretical studies have concentrated on the simultaneous evaluation of both the intrinsic and extrinsic contributions in disordered alloys within the framework of the coherent potential approximation (CPA)\cite{PhysRevLett.105.266604,PhysRevB.84.214436,PhysRevB.86.014405,PhysRevB.89.224422,PhysRevB.92.224421}. 
In these studies, the total $\sigma_{xy}$ in disordered systems is calculated by substituting the single-particle Green's functions obtained from the CPA into that in the Kubo--Bastin\cite{Bastin19711811} or Kubo--Streda\cite{0022-3719-15-22-005} formula.
In addition, there have been attempts to separate the obtained total $\sigma_{xy}$ into the intrinsic and extrinsic parts\cite{PhysRevLett.105.266604,PhysRevB.84.214436,PhysRevB.86.014405}
because in the metallic region, where the impurity scattering can be considered in the perturbative approach, it had been revealed that these three contributions were well separated from the diagrammatic scheme\cite{PhysRevB.64.094434,PhysRevB.75.045315,2016arXiv160403111M}.    
The intrinsic part should contain the interband process so that the correlation function between the velocity operators $v_x$ and $v_y$ remains a finite value even in the absence of disorder.
%The band transition of this inter-band process originates from the inter-band component of $v_{x}$ and $v_{y}$.
%On the other hand, the divergent behavior of the skew-scattering part in the dilute limit indicates the intra-band process, where both $v_{x}$ and $v_{y}$ have intra-band element.
On the other hand, the divergent behavior of the skew-scattering part in the dilute limit originates from the intraband conductive process, where both $v_{x}$ and $v_{y}$ have intraband matrix elements.
In this process, %the antisymmetric part of $\braket{v_xv_y}$ is induced from the electron scattering containing SOI in the form of vertex corrections (VC).
the correlation function of $v_{x}$ and $v_{y}$ needs vertex correction (VC) terms containing the SOI to realize an even parity in terms of $k_{x}$ and $k_{y}$, which are the wave vectors of the electrons.
According to a previous analysis\cite{PhysRevB.64.094434}, the side-jump part is ascribed to the scattering process with VC as well as the skew-scattering part.
From this framework, the intrinsic and extrinsic parts of $\sigma_{xy}$ were defined as the conductive process NOT including and including VC.

However, recent model calculations discussed the side-jump part from the viewpoint of the connection with the Berry phase of electrons using the semiclassical Boltzmann equation\cite{PhysRevB.72.045346,PhysRevB.73.075318}.
This effect is understood as the interference effect between the interband and intraband processes in the correlation functions of $v_x$ and $v_y$\cite{PhysRevB.75.045315}, i.e., one matrix element of either $v_{x}$ or $v_{y}$ is an interband element and the other is an intraband element.
Reflecting this scheme, each part of $\sigma_{xy}$ should be classified by the matrix elements of $v_{x}$ and $v_{y}$ in the Kubo--Bastin formula, as shown in Fig. \ref{fig:fig1}, rather than the vertex corrections in the conductive process.
%rather than the VC in the conductive process.
The leading contribution of the defined intrinsic part is given without VC terms, whereas the skew-scattering part is dominated by the conductive process with the VC terms mentioned above.
In contrast to these two parts, the side-jump term is expected to consist of both processes with and without VC, as shown in previous analyses\cite{PhysRevB.75.045315,PhysRevLett.105.036601}.

In this study, on the basis of the scheme in Fig. \ref{fig:fig1}, we present a first-principles method for evaluating each origin (intrinsic, side-jump, and skew-scattering) of $\sigma_{xy}$ using the tight-binding linear muffin-tin orbital method (TB-LMTO) method with the CPA. %, which rests on the conductive process mentioned in the recent model calculation\cite{PhysRevB.73.075318}, using the Tight-Binding linear muffin-tin orbital method (TB-LMTO) method within CPA.
We also apply our evaluation method to FePt and FePd disordered alloys.
The calculated value of each origin indicates its validity in terms of its physical properties such as the influence on the presence of electron scattering and the dependence on $\sigma_{xx}$ in Eq. (\ref{eq:00}).

This paper is organized as follows. Our calculation method is explained in sec. 2.
Sec. 2 A presents the framework of our first-principles technique to separate each origin of $\sigma_{xy}$. %, and confirms its correspondence with the result of recent perturbation theory\cite{PhysRevB.75.045315}.
In sec. 2 B and 2 C, we respectively explain the effective Hamiltonian of the disordered system and the calculation procedure of $\sigma_{xy}$ from the Kubo--Bastin formula in the TB-LMTO-CPA method.
Sec. 2 D presents the calculation details of each origin of $\sigma_{xy}$ using the TB-LMTO-CPA method based on the schemes in sec. 2 A.
The results of this calculation method and the physical validity of the obtained results are demonstrated for FePt and FePd disordered alloys in sec. 3.
\section{Calculation method}

\subsection{Framework for the distinction of each part of the anomalous Hall conductivity}

\begin{figure}[t]
\includegraphics[width=0.43\textwidth]{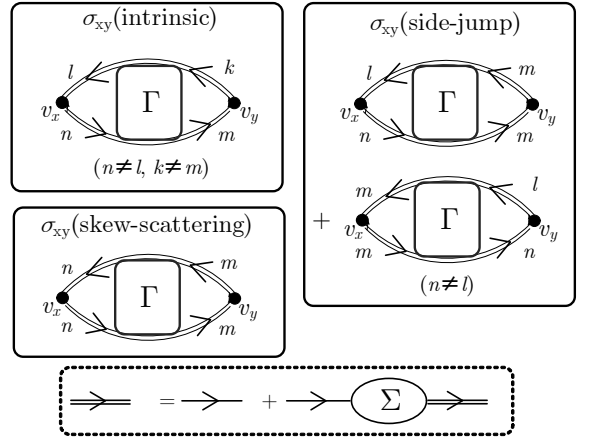}
\caption{Definitions of each part of $\sigma_{xy}$ in this study using a band representation. The band indices are represented by $(l,m,n,k)$, and $v_{\mu}$ is the velocity operator. $\Sigma$ and $\Gamma$ are the self-energy and the vertex correction with respect to the scattering potential, respectively. We distinguish each part of $\sigma_{xy}$ by whether the velocity operators $v_x$ and $v_y$ have diagonal or off-diagonal elements in the band representation. }
\label{fig:fig1}
\end{figure}

The direct electric conductivity (DC) tensor at $0$~K is expressed according to the Kubo--Bastin formula as follows\cite{0022-3719-15-22-005}\cite{Bastin19711811}:
\begin{align}
\sigma_{xy} = -2\sigma_0\int_{-\infty}^{E_{\mathrm{F}}} \mathrm{d}E\mathrm{Tr} & \left\{ V_{\mu}G_+^{'}(E)V_{\nu}\left[G_+(E)-G_-(E)\right] \right. \notag \\
\hspace{1em} -  & \left. V_{\mu}\left[G_+(E)-G_-(E)\right]V_{\nu}G_-^{'}(E)\right\}\label{eq:-1}
\end{align}

where $\mu$ and $\nu$ denote indices in Cartesian coordinates $(x,y,z)$, and $\sigma_{0} = (\hbar e^2)/(4\pi\Omega)$, where $\Omega$ is the volume of the system.
$V_{\mu}$ is the velocity operator of electrons along the $\mu$ axis.
$G_+(E) (G_-(E))$ is the retarded (advanced) Green's functions of a single-particle electron, and $G_{\pm}^{'}(E)$ is their energy derivative.
After the partial integral about the energy\cite{PhysRevB.64.094434}, eq. (\ref{eq:-1}) is divided into the two parts;

\begin{align}
\sigma_{\mu\nu} = & \sigma_0\mathrm{Tr}\left\{V_{\mu}\left(G_+(E_{\mathrm{F}})-G_-(E_{\mathrm{F}})\right)V_{\nu}G_-(E_{\mathrm{F}})\right. \notag \\
&\hspace{1.5em}-\left.V_{\mu}G_+(E_{\mathrm{F}})V_{\nu}\left(G_+(E_{\mathrm{F}})-G_-(E_{\mathrm{F}})\right)\right\}\notag \\
& - \sigma_0 \int_{-\infty}^{E_{\mathrm{F}}}\mathrm{d}E \notag \\
&\times \mathrm{Tr}\left\{ V_{\mu} G_+^{'}(E) V_{\nu}G_+(E) - V_{\mu} G_+(E) V_{\nu} G_+^{'}(E) \right. \notag \\
& \hspace{1.5em} \left. - V_{\mu} G_-^{'}(E) V_{\nu} G_-(E) + V_{\mu} G_-(E) V_{\nu} G_-^{'}(E) \right\}, \label{eq:0}\\
= & \sigma_{\mu\nu}^{(1)}+\sigma_{\mu\nu}^{(2)},\label{eq:1}
\end{align}

%The argument of the energy in Green's functions in Eq. (\ref{eq:1}) is omitted here.
The first term ($\sigma_{\mu\nu}^{(1)}$) is the Fermi-surface term, which represents the contributions of electrons at $E=E_{\mathrm{F}}$. In contrast, the second term ($\sigma_{\mu\nu}^{(2)}$) is the Fermi-sea term, which follows from the contributions $E < E_{\mathrm{F}}$.
$\sigma_{xy}$ is given by sum of the both terms, although $\sigma_{xx}$ consists of only the Fermi-surface term.
%For each origin of the terms in $\sigma_{xy}$,
The Fermi-surface term is regarded as both intrinsic and extrinsic origins because this term has conductive processes originating from electron scattering; these contributions correspond to the extrinsic origin\cite{PhysRevB.75.045315,PhysRevLett.105.036601}.
On the other hand, the Fermi-sea term is regarded as only the intrinsic origin because the scattering effect simply broadens the spectrum of electrons\cite{PhysRevLett.105.036601}.
We classify the various origins of $\sigma_{xy}$ for only the Fermi-surface term in this study.
Fig. \ref{fig:fig1} shows the definition of each origin in the Fermi-surface term.
The double lines and black dots denote the single-particle Green's function involving the self-energy and the velocity operator, respectively.
These pictures are illustrated with a band representation.
The vertex corrections expressed by $\Gamma$ are considered self-consistently by the CPA-vertex correction\cite{PhysRevB.73.144421} in our calculation.
We define the contributions of the three origins of $\sigma_{xy}$ in the Kubo--Bastin formula by distinguishing the intraband and interband elements of $v_x$ and $v_y$ mentioned in the introduction. %, distinct from the previous first-principles calculations . %as shown fig. \ref{fig:fig1}.%; where the conductive process of each origin consists of the groups of $v_x$ and $v_y$ in table 0.
It is easily found that the sum of the three parts in fig. \ref{fig:fig1} is equivalent to the total $\sigma_{xy}^{(1)}$ in Eq. (\ref{eq:1}).

In clean systems, the defined intrinsic part corresponding to the total $\sigma_{xy}$ is given by
\begin{align}
\sigma_{xy} = \frac{2\hbar e^2}{\Omega} \sum_{\bm{k}}\sum_{n,m\neq n} \frac{\mathrm{Im}\left\{\braket{n,\bm{k}|v_x|m,\bm{k}}\braket{m,\bm{k}|v_y|n,\bm{k}}\right\}}{E_{n,\bm{k}}-E_{m,\bm{k}}},\label{eq:-}
\end{align}
where $n$ and $E_{n,\bm{k}}$ are the band index and the eigenenergy of the $(n,\bm{k})$ state, respectively.
One can find that the total $\sigma_{xy}$ in Eq. (\ref{eq:-}) consists of the interband elements of $v_{x}$ and $v_{y}$, which are categorized into the intrinsic part in our definition.
In addition, in sec. 3 we confirm that the intrinsic part slightly changes in the dilute impurity region of FePt and FePd alloys, whereas the two extrinsic parts exhibit noncontiguous behaviors when systems become disordered. 
It should be noted that a recent model calculation\cite{PhysRevB.75.045315} shows two kinds of diagrams for the skew-scattering contributions, which have independent and proportional dependencies on the relaxation time of electrons $\tau$.
Our defined skew-scattering processes involve both contributions, and we combine both contributions as the skew-scattering part in this study.

\subsection{Effective Hamiltonian of disordered systems in the TB-LMTO method}
In pure systems, the TB-LMTO Hamiltonian under the scalar-relativistic approximation with the SOI is expressed by an orthogonal basis as\cite{TurekDrchalKudrnovskySobWeinberger199612,doi:10.1080/14786430802232553}
\begin{align}
H = & C+\Delta^{1/2}S^0\left[1-\gamma S^0\right]^{-1}\Delta^{1/2} \notag \\
= & C + \Delta^{1/2}S^{\gamma}\Delta^{1/2}, \label{eq:2}
\end{align}
%where $\bm{R}$ denotes the position of atoms.
where $C$, $\Delta$, and $\gamma$ are the site-diagonal matrices called ''potential parameters\cite{TurekDrchalKudrnovskySobWeinberger199612}.''
$S^0$ is the structure constant having intersite matrix elements, and $S^{\gamma} = S^0 \left(1-\gamma S^0\right)^{-1}$.
The first term denotes the isolated energy of atoms including the SOI, and the second term is the hopping energy.
The Green's function is introduced by $G(z)=(z-H)^{-1}$, and the specific form of $G(z)$ is given by
\begin{align}
G(z) = \Delta^{-1/2}\left\{\Delta^{-1/2}(z-C)\Delta^{-1/2}-S^{\gamma}\right\}^{-1}\Delta^{-1/2}.\label{eq:3}
\end{align}
For the application of the CPA, this Green's function can be transformed as
\begin{align}
G(z) = \lambda^{\alpha}(z) + \mu^{\alpha}(z)g^{\alpha}(z)\left(\mu^{\alpha}(z)\right)^t,\label{eq:4}
\end{align}
where $g^{\alpha}(z) = (P^{\alpha}(z) - S^{\alpha})^{-1}$ is the so-called auxiliary Green's function. $S^{\alpha} = S^0(1-\alpha S^0)^{-1}$, and $P^{\alpha}(z)$, $\lambda^{\alpha}(z)$, and $\mu^{\alpha}(z)$ are the site-diagonal matrices expressed as
\begin{align}
P^{\alpha}(z) = &\left\{ \Delta^{1/2} (z-C)^{-1} \Delta^{1/2} + (\gamma-\alpha) \right\}^{-1},\label{eq:5.0}\\
\mu^{\alpha}(z) =& \Delta^{-1/2}\left\{ 1 + (\alpha-\gamma) P^{\alpha}(z) \right\}, \label{eq:5}\\
\lambda^{\alpha}(z) = & \mu^{\alpha}(z)(\gamma-\alpha) \Delta^{-1/2} \label{eq:6}.
\end{align}
The CPA is applicable to the Green's function expressed by the representation $\alpha$, which is independent of the constituent atoms.
After applying the CPA, the Green's function $\bar{G}^{\pm}(E)$, where + and - denote the retarded and advanced Green's functions, respectively, is given by the site representation as
\begin{align}
&\bar{G}^{\pm}_{\bm{R},\bm{R}'}(E)  \notag \\=  & \delta_{\bm{R},\bm{R}'} \sum_{Q} c_{\bm{R}}^{Q}\notag \\
&\times \left\{ \lambda_{\bm{R}}^{\alpha,Q}(E) + \mu_{\bm{R}}^{\alpha,Q}(E) \bar{g}_{\bm{R},\bm{R}}^{\alpha,\pm}(E) f_{\bm{R}}^{\alpha,Q\pm}(E) \left(\mu_{\bm{R}}^{\alpha,Q}(E)\right)^t \right\}  \notag \\
& + \left(1-\delta_{\bm{R},\bm{R}'}\right) \sum_{Q,Q'} c_{\bm{R}}^Qc_{\bm{R}'}^{Q'}\notag \\
& \times \mu_{\bm{R}}^{\alpha,Q}(E)  \tilde{f}_{\bm{R}}^{\alpha,Q,\pm}(E) \bar{g}^{\alpha,\pm}_{\bm{R},\bm{R}'}(E) f_{\bm{R}'}^{\alpha,Q',\pm}(E) \left(\mu_{\bm{R}'}^{\alpha,Q'}(E)\right)^t , \label{eq:7}
\end{align}
where $c_{\bm{R}}^Q$ is the occupation ratio of the $Q$ atom at the $\bm{R}$ site, and $\lambda_{\bm{R}}^Q(E)$ and $\mu_{\bm{R}}^Q(E)$ are the corresponding values of the $Q$ atom.
$\bar{g}^{\alpha,\pm}(E) = (\bar{P}^{\alpha,\pm}(E)-S^{\alpha})^{-1}$, and $\bar{P}^{\alpha,\pm}(E)$ is the coherent potential function, which is self-consistently determined by the CPA.
%$f_{\bm{R}}^{\alpha,Q,\pm}$ and $\tilde{f}_{\bm{R}}^{\alpha,Q,\pm}$ fulfill the relations;
$f_{\bm{R}}^{\alpha,Q,\pm}(E)$ is obtained from the CPA conditions\cite{TurekDrchalKudrnovskySobWeinberger199612}:
\begin{align}
%\tilde{f}_{\bm{R}}^{\alpha,Q,\pm} \bar{g}_{\bm{R},\bm{R}}^{\alpha,\pm} = \bar{g}_{\bm{R},\bm{R}}^{\alpha,\pm} f_{\bm{R}}^{\alpha,Q,\pm} = g_{\bm{R},\bm{R}}^{\alpha,Q} = (P^{\alpha,Q}-S^{\alpha})_{\bm{R},\bm{R}}^{-1}.\label{eq:8}
f_{\bm{R}}^{\alpha,Q,\pm}(E) = \left\{ 1 + \left(P_{\bm{R}}^{\alpha,Q}(E) - \bar{P}_{\bm{R}}^{\alpha,\pm}(E)\right) \bar{g}_{\bm{R},\bm{R}}^{\alpha,\pm}(E) \right\}^{-1}, \label{eq:8}
\end{align}
and $\tilde{f}_{\bm{R}}^{\alpha,Q,\pm}(E)$ is the transposed matrix of ${f}_{\bm{R}}^{\alpha,Q,\pm}(E)$. The effective Hamiltonian $H_{\mathrm{eff}}^{\pm(E)}$ in the disordered system is introduced as\cite{PhysRevB.41.7515}
\begin{align}
H_{\mathrm{eff}}^{\pm}(E) = E \pm i \delta -\left(\bar{G}^{\pm}(E)\right)^{-1},\label{eq:8.1}
\end{align}
where $\delta$ is an infinitesimal number that gives the retarded and advanced Green's functions before the CPA as $G^{\pm}(E) = G(E\pm i\delta)$. $H_{\mathrm{eff}}^{\pm}(E)$ is rewritten in a similar form to Eq. (\ref{eq:2}) as
\begin{align}
%H_{\mathrm{eff}}^{\pm}(E) = \bar{C}_{\pm}(E) + \left(\bar{\Delta}_{\pm}^t(E)\right)^{1/2} \left[S^{\alpha}\left(1+\left\{\alpha-\bar{\gamma}_{\pm}(E)\right\}S^{\alpha}\right)^{-1}\right] \left(\bar{\Delta}_{\pm}(E)\right)^{1/2},\label{eq:9}
&H_{\mathrm{eff}}^{\pm}(E) = \bar{C}_{\pm}(E) + \left(\bar{\Delta}_{\pm}(E)\right)^{1/2} \notag \\
& \hspace{3.5em} \times \left[S^{\alpha}\left(1+\left\{\alpha-\bar{\gamma}_{\pm}(E)\right\}S^{\alpha}\right)^{-1}\right] \left(\bar{\Delta}_{\pm}(E)\right)^{1/2},\label{eq:9}
\end{align}
where $\bar{C}_{\pm}(E)$, $\bar{\Delta}_{\pm}(E)$, and $\bar{\gamma}_{\pm}(E)$ are the quantities corresponding to $C$, $\Delta$, and $\gamma$ in ordered alloys in Eq. (\ref{eq:2}), respectively.
($\bar{C}_{\pm}(E)$, $\bar{\Delta}_{\pm}(E)$, $\bar{\gamma}_{\pm}(E)$) are site-diagonal as well as ($C$, $\Delta$, $\gamma$), but they have complex elements that satisfy ($\bar{C}_+(E)$, $\bar{\Delta}_+(E)$, $\bar{\gamma}_+(E)$) = ($\bar{C}_-(E)$, $\bar{\Delta}_-(E)$, $\bar{\gamma}_-(E)$)$^*$, which originates from the imaginary parts of the energy; these complex elements are different from ($C$, $\Delta$, $\gamma$), which consists of real numbers.
Then, the effective Hamiltonian of the disordered system in Eq. (\ref{eq:9}) has anti-Hermitian elements that are different from those of the ordered system.

\subsection{Calculation method of the AHE from the Kubo--Bastin formula in the TB-LMTO method}
%In this section, we transform $\sigma_{xy}$ in Eq. (\ref{eq:0}) into a more convenient form for the TB-LMTO-CPA method according to previous studies\cite{PhysRevB.86.014405,PhysRevB.89.064405}.
In the TB-LMTO approach, the DC conductivity can be calculated by applying the Hamiltonian of the TB-LMTO method in Eq. (\ref{eq:2}) to the Kubo--Bastin formula in Eq. (\ref{eq:0})\cite{PhysRevB.65.125101,PhysRevB.86.014405}.
The velocity operator along $\mu$-axes in Eq. (\ref{eq:0}) is given by the commutation relation as follows\cite{PhysRevB.86.014405}:
\begin{align}
V_{\mu} = \left(i\hbar\right)^{-1}\left[X_{\mu},H\right],\label{eq:10}
\end{align}
where $H$ is the Hamiltonian of the TB-LMTO method, and $X_{\mu}$ is the position operator, which is diagonal about the site position $\bm{R}$ and the orbital index of electrons $L$ as
\begin{align}
(X_{\mu})_{\bm{R},\bm{R}',L,L'} = X_{\mu}^{\bm{R}}\delta_{\bm{R},\bm{R}'}\delta_{L,L'}.\label{eq:11}
\end{align}
Consequently, the practical form of $V_{\mu}$ in the TB-LMTO approach is given by
\begin{align}
V_{\mu} = \Delta^{1/2}\{1+S^{\alpha}(\alpha-\gamma)\}^{-1}v_{\mu}^{\alpha}\{1+(\alpha-\gamma)S^{\alpha}\}^{-1}\Delta^{1/2},\label{eq:12}
\end{align}
where $v_{\mu}^{\alpha}$ is the following velocity-like quantity\cite{PhysRevB.86.014405};
\begin{align}
v_{\mu}^{\alpha} = (i\hbar)^{-1}\left[X_{\mu},S^{\alpha}\right].\label{eq:13}
\end{align}
For deriving Eq. (\ref{eq:12}), we utilize two relations. First, the commutativity between $X_{\mu}$ and the site-diagonal matrix $\Delta^{1/2}$ is used, and second, the following commutation relation is applied:
\begin{align}
&\left[X_{\mu} , S^{\alpha} \left(1+(\alpha-\gamma)S^{\alpha}\right)^{-1}\right] \notag \\
= &\{1+S^{\alpha}(\alpha-\gamma)\}^{-1}v_{\mu}^{\alpha}\{1+(\alpha-\gamma)S^{\alpha}\}^{-1}.\label{eq:14}
\end{align}
%In addition, $G(z)$ in Eq. () can be rewritten based on Eq. ( and ) and $g^{\alpha}(z) = \left(P^{\alpha}(z)-S^{\alpha}\right)^{-1}$ as
%\begin{align}
%G(z) = 
%\end{align}
Substituting Eq. (\ref{eq:4}, \ref{eq:5}, and \ref{eq:6}) and Eq. (\ref{eq:12}) into Eq. (\ref{eq:0}), $\sigma_{\mu\nu}^{(1)}$ and $\sigma_{\mu\nu}^{(2)}$ in Eq. (\ref{eq:1}) are respectively transformed as follows:
\begin{align}
&\sigma_{\mu\nu}^{(1)}\notag \\
 = & \sigma_0 \mathrm{Tr} \left. \left\{v_{\mu}^{\alpha} \left(g_+^{\alpha}(E_{\mathrm{F}}) -g_-^{\alpha}(E_{\mathrm{F}})\right) v_{\nu}g_-^{\alpha}(E_{\mathrm{F}})\right.\right.\notag \\
& \hspace{2em} \left.\left. - v_{\mu}^{\alpha} g_+^{\alpha}(E_{\mathrm{F}}) v_{\nu}^{\alpha} \left(g_+^{\alpha}(E_{\mathrm{F}})-g_-^{\alpha}(E_{\mathrm{F}}) \right)\right\} \right.\notag \\
%\sigma_{\mu\nu}^{(1)} = & \frac{\hbar e^2}{4\pi\Omega} \mathrm{Tr} \left.\braket{v_{\mu
& + \sigma_{0}\mathrm{Tr}\Bigl\{(\alpha-\gamma)\left\{1+S^{\alpha}(\alpha-\gamma)\right\}^{-1} \notag \\
& \hspace{2em} \times \left[v_{\mu}^{\alpha}(g_+^{\alpha}(E_{\mathrm{F}})-g_-^{\alpha}(E_{\mathrm{F}}))v_{\nu}^{\alpha} \right. \notag \\
& \hspace{2.5em} \left. - v_{\nu}^{\alpha}(g_+^{\alpha}(E_{\mathrm{F}})-g_-^{\alpha}(E_{\mathrm{F}}))v_{\mu}^{\alpha}\right]\Bigr\}, \label{eq:15}%\right|_{E=E_{\mathrm{F}}}, \label{eq:15}
\end{align}
%\vspace{-0.7cm}
\begin{align}
 \sigma_{\mu\nu}^{(2)}
= & - \sigma_0\int_{-\infty}^{E_{\mathrm{F}}}\mathrm{d}E \notag \\
&\times \mathrm{Tr}\left\{v_{\mu}(g_+^{\alpha}(E))'v_{\nu}^{\alpha}g_+(E)-v_{\mu}^{\alpha}g_+^{\alpha}(E)v_{\nu}^{\alpha}(g_+^{\alpha}(E))'\right.\notag \\
&\hspace{2em}-\left.v_{\mu}(g_-^{\alpha}(E))'v_{\nu}^{\alpha}g_-(E)-v_{\mu}^{\alpha}g_-^{\alpha}(E)v_{\nu}^{\alpha}(g_-^{\alpha}(E))' \right\}\notag \\
 &-  \sigma_{0}\int_{-\infty}^{E_{\mathrm{F}}}\mathrm{d}E \mathrm{Tr}{X_{\mu\nu}}(E), \label{eq:16}
\end{align}
%\vspace{-0.5cm}
\begin{align}
X_{\mu\nu}(E)= & (\alpha-\gamma)\left\{1+S^{\alpha}(\alpha-\gamma)\right\}^{-1}\notag \\
&\times \left[v_{\mu}^{\alpha}\left\{(g_+^{\alpha}(E))'-(g_-^{\alpha}(E))'\right\}v_{\nu}^{\alpha}\right. \notag \\
&\hspace{0.5em}\left.-v_{\nu}^{\alpha}\left\{(g_+^{\alpha}(E))'-(g_-^{\alpha}(E))'\right\}v_{\mu}^{\alpha}\right].\label{eq:17}
%- & \left\{1+S^{\alpha}(\alpha-\gamma)\right\}^{-1}(\alpha-\gamma). 
\end{align}
We can simplify the total $\sigma_{xy}$, which is the sum of Eq. (\ref{eq:15}) and Eq. (\ref{eq:16})
by integrating the second term of Eq. (\ref{eq:16}); this integration result cancels out the second term in Eq. (\ref{eq:15})\cite{PhysRevB.86.014405}. % because the second term of Eq. (\ref{eq:16}) has energy arguments only in $g_{\pm}^{\alpha}$.
As a result, the total $\sigma_{xy}$ is rewritten as
\begin{align}
&\sigma_{\mu\nu} \notag \\
 = & \sigma_0 \mathrm{Tr} \left. \left\{v_{\mu}^{\alpha} \left(g_+^{\alpha}(E_{\mathrm{F}}) -g_-^{\alpha}(E_{\mathrm{F}})\right) v_{\nu}g_-^{\alpha}(E_{\mathrm{F}})\right.\right.\notag \\
& \hspace{2em} \left.\left. - v_{\mu}^{\alpha} g_+^{\alpha}(E_{\mathrm{F}}) v_{\nu}^{\alpha} \left(g_+^{\alpha}(E_{\mathrm{F}})-g_-^{\alpha}(E_{\mathrm{F}}) \right)\right\} \right.\notag \\
& - \sigma_0\int_{-\infty}^{E_{\mathrm{F}}}\mathrm{d}E\notag \\
&\times \mathrm{Tr}\left\{v_{\mu}(g_+^{\alpha}(E))'v_{\nu}^{\alpha}g_+^{\alpha}(E)-v_{\mu}^{\alpha}g_+^{\alpha}(E)v_{\nu}^{\alpha}(g_+^{\alpha}(E))'\right.\notag \\
&\hspace{1.5em}-\left.v_{\mu}(g_-^{\alpha}(E))'v_{\nu}^{\alpha}g_-^{\alpha}(E)+v_{\mu}^{\alpha}g_-^{\alpha}(E)v_{\nu}^{\alpha}(g_-^{\alpha}(E))'\right\}. \label{eq:18}
\end{align}
This transformed $\sigma_{xy}$ involves the representation $\alpha$, which does not exist in the original form in Eq. (\ref{eq:0}).
However, Eq. (\ref{eq:18}) is shown to be independent of this representation\cite{PhysRevB.89.064405}.
Rather, Eq. (\ref{eq:18}) has advantages with regard to the application of the CPA in disordered alloys compared with Eq. (\ref{eq:0}); when $\alpha$ is independent of the atomic species, $v_{\mu}^{\alpha}$ is also independent of them. %does NOT need configuration average different from $V_{\mu}$ when $\alpha$ is independent of atomic species. %ecause $v_{\mu}^{\alpha}$ is NOT a random quantity for atomic species different from $V_{\mu}$.
A more concrete treatment for calculating Eq. (\ref{eq:18}) has been discussed in previous studies\cite{PhysRevB.86.014405,PhysRevB.89.064405,PhysRevB.89.064405}.

\subsection{Evaluation method of each part of the AHE from the Kubo--Bastin formula}
%$\braket{\sigma_{xy}^{(1,2)}}_{\mathrm{conf}}$
In this section, a practical procedure for separating $\sigma_{xy}$ into each origin on the basis of the concepts in fig. \ref{fig:fig1} is presented.
We address this separation only in $\sigma_{xy}^{(1)}$ in Eq. (\ref{eq:1}), whereas we regard all of the contributions of $\sigma_{xy}^{(2)}$ as the intrinsic part, as mentioned in sec. 2A.
We distinguish each origin of $\sigma_{xy}^{(1)}$ by the matrix elements %intra- or inter-band part
of $V_x$ and $V_y$ in the band representation of $\sigma_{xy}^{(1)}$.
The target of this evaluation is disordered alloys having both intrinsic and extrinsic origins of $\sigma_{xy}$.
One difficulty for performing this division method in disordered systems is that the band representation depends on the atoms occupying the sites.
The unitary matrix $U$, which converts the TB-LMTO Hamiltonian into a band representation, is introduced to satisfy the following relationship:
\begin{align}
\left\{ U^{\dagger}HU \right\}_{n,m} = \varepsilon_{n} \delta_{n,m},\label{eq:19}
\end{align}
%where $Q, Q'$ is the occupied atom, which is random in the disordered system, and $n,m$ are the band indices.
% $H_{Q}$ is the Hamiltonian of $Q$-atom given in Eq. (\ref{eq:2}), and $\varepsilon_{n,Q,Q'}$ is the eigenenergy of $n,Q,Q'$ states.
where $n$ is the eigenstate, $\varepsilon_n$ is the eigenenergy of state $n$, %the in the disordered system.
and $H$ is the Hamiltonian given by Eq. ({\ref{eq:2}). These quantities are uncertain because they depend on the randomly located atoms in disordered systems.
To resolve this issue, we replace $H$ with the nonrandom effective Hamiltonian $H^{\pm}_{\mathrm{eff}}$ in Eq. (\ref{eq:9}) based on the CPA and define the band representation as follows:
\begin{align}
& \left\{U^{\dagger}(E)\left[\frac{H_{\mathrm{eff}}^{+}(E)+H_{\mathrm{eff}}^{-}(E)}{2}\right]U(E)\right\}_{\bm{k},n,m} \notag \\
= & \left(\varepsilon_{\mathrm{eff}}\right)_{\bm{k},n} \delta_{n,m}.\label{eq:20}
\end{align}
The actual diagonalized object is the Hermitian part of $H_{\mathrm{eff}}^{\pm}(E)$ because the original $H_{\mathrm{eff}}^{\pm}(E)$ in Eq. (\ref{eq:9}) involves the anti-Hermitian part due to the scattering effect of electrons.
We used $ H_{\mathrm{eff}}^-(E) = \{H_{\mathrm{eff}}^+(E)\}^{\dagger}$ to obtain the Hermitian conjugate.
The corresponding effective velocity operator is introduced as
\begin{align}
\left(V_{\mathrm{eff}}^{\pm}(E)\right)_{\mu} = (i\hbar)^{-1}\left[X_{\mu},H_{\mathrm{eff}}^{\pm}(E)\right].\label{eq:21}
\end{align}
Via a similar transformation to that of Eq. (\ref{eq:12}), Eq. (\ref{eq:21}) is rewritten as
\begin{align}
\left(V_{\mathrm{eff}}^{\pm}(E)\right)_{\mu} = & \left(\bar{\Delta}_{\pm}(E)\right)^{1/2}\{1+S^{\alpha}(\alpha-\bar{\gamma}_{\pm}(E))\}^{-1}v_{\mu}^{\alpha}\notag \\
\times & \{1+(\alpha-\bar{\gamma}_{\pm}(E))S^{\alpha}\}^{-1}(\bar{\Delta}_{\pm}(E))^{1/2}.\label{eq:22}
\end{align}
We define the diagonal (nondiagonal) part of $\left(V_{\mathrm{eff}}^{\pm}(E)\right)_{\mu}$, which is important for the separation of each origin of $\sigma_{xy}$ according to fig. \ref{fig:fig1}, as $\left(V_{\mathrm{eff}}^{\mathrm{d},\pm}(E)\right)_{\mu}$ $\left(\left(V_{\mathrm{eff}}^{\mathrm{nd},\pm}(E)\right)_{\mu}\right)$, given by
\begin{align}
&\left\{U^{\dagger}(E)\left(V_{\mathrm{eff}}^{\mathrm{d},\pm}(E)\right)_{\mu}U(E)\right\}_{n,m} \notag \\
 = &\left\{U^{\dagger}(E)\left(V_{\mathrm{eff}}^{\pm}(E)\right)_{\mu}U(E) \right\}_{n,n}\delta_{n,m},\label{eq:22.1}\\
&\left\{U^{\dagger}(E)\left(V_{\mathrm{eff}}^{\mathrm{nd},\pm}(E)\right)_{\mu}U(E)\right\}_{n,m} \notag \\
= &\left\{U^{\dagger}(E)\left(V_{\mathrm{eff}}^{\pm}(E)\right)_{\mu}U(E) \right\}_{n,m}\left(1-\delta_{n,m}\right).\label{eq:23}
%\left\{U^{\dagger}\left(V_{\mathrm{eff}}^{\mathrm{d},\pm}\right)_{\mu}U\right\}_{n,m} = \left\{U^{\dagger}\left(V_{\mathrm{eff}^{\pm}\right)_{\mu}\right\}_{n,n}\delta_{n,m}
\end{align}
From the above definitions, $\left(V_{\mathrm{eff}}^{\mathrm{d(nd)},\pm}(E)\right)_{\mu}$ satisfies the following relation:
\begin{align}
\left(V_{\mathrm{eff}}^{\mathrm{d},\pm}(E)\right)_{\mu} + \left(V_{\mathrm{eff}}^{\mathrm{nd},\pm}(E)\right)_{\mu} = \left(V_{\mathrm{eff}}^{\pm}(E)\right)_{\mu}.\label{eq:24}
\end{align}
In the practical calculation of each part of $\sigma_{xy}$, we do not use Eq. (\ref{eq:0}) but instead use Eq. (\ref{eq:15} and \ref{eq:16}) to perform the configuration average within the framework of the CPA\cite{PhysRevB.86.014405,PhysRevB.89.064405}. %, which consists of $v_{\mu}^{\alpha}$, because
We introduce $\left(v_{\mathrm{eff}}^{\mathrm{d(nd)},\pm}(E)\right)_{\mu}$, which is the diagonal (nondiagonal) part of $v_{\mu}$ in the band representation, from Eq. (\ref{eq:22}) as %and the correspondence relation with $\left(V_{\mathrm{eff}}^{\mathrm{d(nd)},\pm}\right)_{\mu}$ as; % defined in Eq. ()\left(v_{\mathrm{eff}}^{\mathrm{d(nd)},\pm}\right)_{\mu} which is similar
\begin{align}
&\left(V_{\mathrm{eff}}^{\mathrm{d(nd)}\pm}(E)\right) \notag \\
= & \left(\bar{\Delta}_{\pm}(E)\right)^{1/2}\{1+S^{\alpha}(\alpha-\bar{\gamma}_{\pm}(E))\}^{-1}\left(v_{\mathrm{eff}}^{\mathrm{d(nd)},\pm}(E)\right)_{\mu}^{\alpha}\notag \\
& \times \{1+(\alpha-\bar{\gamma}_{\pm}(E))S^{\alpha}\}^{-1}(\bar{\Delta}_{\pm}(E))^{1/2}.\label{eq:25}
\end{align}
One can show the relation for $\left(v_{\mathrm{eff}}^{\mathrm{d(nd)},\pm}(E)\right)_{\mu}^{\alpha}$ by substituting Eq. (\ref{eq:22}) and Eq. (\ref{eq:25}) into Eq. (\ref{eq:24}):
\begin{align}
\left(v_{\mathrm{eff}}^{\mathrm{d},\pm}(E)\right)_{\mu}^{\alpha}+\left(v_{\mathrm{eff}}^{\mathrm{nd},\pm}(E)\right)_{\mu}^{\alpha} = v_{\mu}^{\alpha}. \label{eq:25.0}
\end{align}
We regard the Hermitian part of $\left(v_{\mathrm{eff}}^{\mathrm{d(nd)},\pm}(E)\right)_{\mu}^{\alpha}$ as the effective velocity as
\begin{align}
\left(v_{\mathrm{eff}}^{\mathrm{d(nd)}}(E)\right)_{\mu}^{\alpha} = \frac{\left(v_{\mathrm{eff}}^{\mathrm{d(nd)},+}(E)\right)_{\mu}^{\alpha}+\left(v_{\mathrm{eff}}^{\mathrm{d(nd)},-}(E)\right)_{\mu}^{\alpha}}{2}. \label{eq:25.1}
\end{align}
From Eq. (\ref{eq:25.0}) and Eq. (\ref{eq:25.1}), the following relationship is given as
%One can show the following relation for $\left(v_{\mu}^{\mathrm{d(nd)}}(E)\right)_{\mu}^{\alpha}$ using Eq. (\ref{eq:24}), and (\ref{eq:25}):
\begin{align}
\left(v_{\mathrm{eff}}^{\mathrm{d}}(E)\right)_{\mu}^{\alpha} + \left(v_{\mathrm{eff}}^{\mathrm{nd}}(E)\right)_{\mu}^{\alpha} = v_{\mu}.\label{eq:26}
\end{align}
Our separation method for each origin of $\sigma_{xy}$ is performed by substituting $\left(v_{\mathrm{eff}}^{\mathrm{d(nd)}}\right)^{\alpha}_{x,y}$ for $v_{x,y}$ in Eq. (\ref{eq:15}).
The separation of each contribution is performed as follows:
\begin{align}
\sigma_{xy}^{\mathrm{int}} =& \sigma_{xy}^{(1)}\left(v_x^{\alpha},v_y^{\alpha}\rightarrow \left(v_{\mathrm{eff}}^{\mathrm{nd}}(E_{\mathrm{F}})\right)_{x}^{\alpha},\left(v_{\mathrm{eff}}^{\mathrm{nd}}(E_{\mathrm{F}})\right)_{y}^{\alpha}\right) + \sigma_{xy}^{(2)}, \label{eq:27}\\
\sigma_{xy}^{\mathrm{sj}} = &\sigma_{xy}^{(1)}\left(v_x^{\alpha},v_y^{\alpha}\rightarrow \left(v_{\mathrm{eff}}^{\mathrm{d}}(E_{\mathrm{F}})\right)_{x}^{\alpha},\left(v_{\mathrm{eff}}^{\mathrm{nd}}(E_{\mathrm{F}})\right)_{y}^{\alpha}\right)\notag \\
&+ \sigma_{xy}^{(1)}\left(v_x^{\alpha},v_y^{\alpha}\rightarrow \left(v_{\mathrm{eff}}^{\mathrm{nd}}(E_{\mathrm{F}})\right)_{x}^{\alpha},\left(v_{\mathrm{eff}}^{\mathrm{d}}(E_{\mathrm{F}})\right)_{y}^{\alpha}\right), \label{eq:28}\\
\sigma_{xy}^{\mathrm{sk}} = &\sigma_{xy}^{(1)}\left(v_x^{\alpha},v_y^{\alpha}\rightarrow \left(v_{\mathrm{eff}}^{\mathrm{d}}(E_{\mathrm{F}})\right)_{x}^{\alpha},\left(v_{\mathrm{eff}}^{\mathrm{d}}(E_{\mathrm{F}})\right)_{y}^{\alpha}\right). \label{eq:29}
\end{align}
The superscripts (int, sj, and sk) indicate (intrinsic, side-jump, and skew-scattering).
The specific form of Eq. (\ref{eq:27})--(\ref{eq:29}) %$\sigma_{xy}^{(1,2)}\left(v_{\mathrm{eff}}^{\mathrm{nd},\alpha}\right)_{x},\left(v_{\mathrm{eff}}^{\mathrm{nd},\alpha}\right)_{y}$ is denoted in the Appendix.
in this study is presented in the Appendix A, and the invariance properties of these equations with respect to the representation $\alpha$ is shown in Appendix B. 
It is easy to show that sum of each origin in Eq. (\ref{eq:27})--(\ref{eq:29}) is equal to the total $\sigma_{xy}$. %in Eq. ().
%$\sigma_{xy}^{(1,2)}\left(v_{\mathrm{eff}}^{\mathrm{nd},\alpha}\right)_{x},\left(v_{\mathrm{eff}}^{\mathrm{nd},\alpha}\right)_{y}$
%Eq. (\ref{eq:27})-(\ref{eq:29}) contain the random quantities, $g_{\pm}$ and $\gamma$.
%In the configuration average, one should consider the vertex corrections between them. % many-body scattering effect called vertex corrections.
%We take into account the vertex correction between $g_{\pm}$\cite{PhysRevB.86.014405} because this effect is necessary to deal with some parts of the leading terms of the extrinsic origin of $\sigma_{xy}$. %in fig. \ref{fig:fig2}; these terms contain vertex corrections between Green functions.
%In contrast, the vertex correction between $g_{\pm}(E)$ and $\gamma$ is disregarded in our calculation. %assuming that this correction gives small contribution for each part of $\sigma_{xy}$.
%It should be noted that the simplified transform of $\sigma_{xy}$ in Eq. (\ref{eq:18}) is inapplicable for each part of $\sigma_{xy}$ in Eq. ().

\section{Numerical calculation}

\begin{figure}[t]
\includegraphics[width=0.48\textwidth]{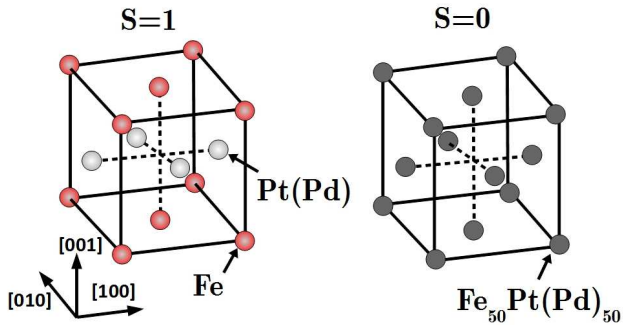}
\caption{Crystal structure and the atom positions of $L\mathrm{1}_{0}$-FePt and FePd alloys having different order parameters $S$.}
\label{fig:fig3}
\end{figure}

In this section, %the applied results of the evaluation method for each origin in $\sigma_{xy}$ are exhibited in $L_{10}$-FePt and FePd alloys.
the calculation of each origin of $\sigma_{xy}$ based on Eq. (\ref{eq:27})--(\ref{eq:29}) is presented for $L\mathrm{1}_{0}$-FePt and FePd alloys.
From the following results, we confirm the validity of the calculated parts of $\sigma_{xy}$ in terms of the dependence of the degree of order in these alloys.
%In addition, we check that the calculated each part of $\sigma_{xy}$ satisfies general relationship between $\sigma_{xx} $Eq.(\ref{eq:0}), which presents the general relationship between the each part and $\sigma_{xx}$.
%In addition, the evaluated value in diluted alloys are shown that rests on perturbation theory are also shown.
\subsection{Implementation conditions}
The employed first-principles technique is the TB-LMTO method under the local spin-density approximation (LSDA).  %\cite{doi:10.1080/14786430802232553}
In addition to the nonrelativistic Hamiltonian, the SOI term in the form of $\xi\bm{l}\cdot\bm{s}$ is introduced. %, which is obtained from the Pauli approximation
The crystal structures of the FePt and FePd alloys are the $L\mathrm{1}_0$-structure, as shown in Fig. \ref{fig:fig3}.
Both structures have four sites in the unit cell, and the average number of valence electrons in the unit cell of these alloys is 36 in common. %, which means similar band structure between these alloys.
%To control the scattering ratio of electrons, we changed randomness of the alignment of Fe and Pt(Pd) atoms at each site, keeping the composition ratio of Fe:Pt(Pd) = 50:50.
To control the scattering ratio of electrons, we introduce chemical disorder and change the degree of order of Fe and Pt(Pd) atoms at each site, keeping the composition ratio of Fe:Pt(Pd) = 50:50.
The probability of existence of the two atoms at each site is described by the order parameter $S=1-2x$ $\left(0\le x\le0.5\right)$, where the occupation ratios at each site are $1-x$:$x$ and $x$:$1-x$.
The relaxation time of electrons in each orbital state is self-consistently given as a function of $S$ using the CPA.
In the ordered alloys, we set the infinitesimal imaginary part $\delta=\pm10^{-5}\left(\mathrm{mRy}\right)$ in the retarded and advanced Green's functions for the numerical calculation of the eigenstates from the singularity point of the Green's functions. %find the singularity point of the Green functions from the numerical calculation.
We employ the equivalent lattice constant along $a$ and $c$ axes of both alloys over the whole range of $S$.
%We fix the ratio of the lattice constants between a and c axes $(c/a)$ of both alloys as $1$ in the whole range of $S$.
%For evaluating $\sigma_{xy}^{(1)}$ in Eq. ([eq:2.26]), the used  $\sim 3\times10^{7}$ $\bm{k}$-points in the Brillouine zone.
In the calculation of $\sigma_{xy}^{(2)}$ in Eq. (\ref{eq:16}), we extend the energy integration along the real axis into a contour integration in the complex plane, where the integral path is like a square, to decrease the necessary integration points for convergence\cite{PhysRevB.26.5433,PhysRevB.89.064405}.
We assume 40 nodes in the upper and lower planes for the integration of the energy in $\sigma_{xy}^{(2)}$.%, and we used $\sim10^{6}$ $\bm{k}$-points in the low energy region and $\sim10^{7}$ $\bm{k}$-points near the Fermi energy in the calculation of each energy node of $\sigma_{xy}^{(2)}$

\begin{figure}[t]
\includegraphics[width=0.4\textwidth]{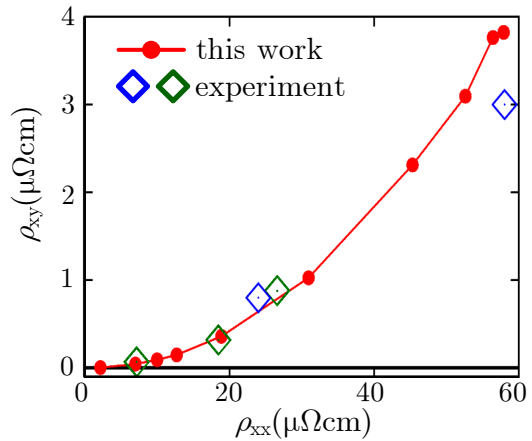}
\caption{The relation between the longitudinal resistivity $\rho_{xx}$ and the anomalous Hall resistivity $\rho_{xy}$ of FePt alloys. The solid line indicates the results obtained from our calculation, and experimental results at low temperature are denoted by the blue\cite{:/content/aip/journal/apl/98/8/10.1063/1.3556616} and green\cite{0022-3727-41-13-135001} rhombuses. Both calculated and experimental results are captured by changing the order parameter $S$ in FePt.}
\label{fig:fig3.1}
\end{figure}

\subsection{Results and discussion}
First, we compare the calculated total $\sigma_{xy}$ of $L\mathrm{1}_{0}$-FePt with the values obtained in the experiment\cite{:/content/aip/journal/apl/98/8/10.1063/1.3556616,0022-3727-41-13-135001} in fig. \ref{fig:fig3.1}.
This graph shows the relation between the longitudinal resistivity $\rho_{xx}$ and the anomalous Hall resistivity $\rho_{xy}$ in these studies. 
The experimental results were obtained by changing the order parameter $S$ of the FePt alloy at a low temperature; these conditions were the same as those in our study.
One can recognize the good agreement between the calculated and experimental results over a wide range of $S$.
Consequently, we confirmed that our calculated total $\sigma_{xy}$ attains reasonable values. %, and go forward to evaluate the result of each calculated origin of $\sigma_{xy}$.

\begin{table}[t]
\caption{Calculated results for each part of $\sigma_{xy}$ in units of $\Omega^{-1}$ cm$^{-1}$ for $S=1$ and $S=0.99$, which respectively indicate the ordered and disordered phases, in FePt and FePd alloys. The results from a previous study assuming the dilute impurity limit are also listed.}
\label{fig:tab1}
\begin{threeparttable}
\begin{tabular}[t]{ccc|c}\hline \hline
FePt & \hspace{0.5em}$S=1$\hspace{0.5em} & $S=0.99$ \hspace{0.5em} &\hspace{0.5em} perturbation\tnote{a} \hspace{0.5em} \\ \hline
$\sigma_{xy}^{\mathrm{int}}$ & 960(289) & 938(250) & 818 \\
$\sigma_{xy}^{\mathrm{sj}}$ & 14 & 126 & 128 \\
$\sigma_{xy}^{\mathrm{sk}}$ & 0 & -562 &  \\ \hline \hline
\end{tabular}

\begin{tabular}[t]{ccc|c}\hline \hline
FePd & \hspace{0.5em}$S=1$\hspace{0.5em} & $S=0.99$ \hspace{0.5em} & \hspace{0.5em}perturbation\tnote{a} \hspace{0.5em}\\ \hline
$\sigma_{xy}^{\mathrm{int}}$ & 173(55) & 171(57) & 133 \\
$\sigma_{xy}^{\mathrm{sj}}$ & 10 & 448 & 263 \\
$\sigma_{xy}^{\mathrm{sk}}$ & 0 & 630 &  \\ \hline \hline
\end{tabular}
\begin{tablenotes}\footnotesize
\item[a]Reference\cite{PhysRevLett.107.106601}.
\end{tablenotes}
\end{threeparttable}
\end{table}

Next, we assess the dependence of each calculated part of $\sigma_{xy}$ in the presence of electron scattering by the disordered system.
Table 1 lists each evaluated value when $S=1$ and $0.99$, which represent the pure and slightly disordered systems, respectively. 
In $\sigma_{xy}^{\mathrm{int}}$, which comprises both Fermi-surface and Fermi-sea terms as shown in Eq. (\ref{eq:27}), we list the Fermi-sea part in the parenthesis in addition to the total value. 
The Fermi-sea parts of both two alloys have almost 30\% contributions of the entire intrinsic value and should NOT disregard as well as the some Heusler alloys\cite{PhysRevB.88.014422}. 
As for the dependence of impurity scattering, one can find that both parts of the intrinsic origin have finite values in ordered ($S=1$) and disordered ($S=0.99$) systems and show similar values between these environments in both alloys. %exhibits a continuous change from $S = 1$ to $S < 1$ in both alloys.
On the other hand, the estimated values of the extrinsic origins (side-jump and skew-scattering) are almost 0 at $S = 1$, and they rapidly increase when the systems become random in both alloys.
These distinct behaviors are explained by the factors of these origins.
The intrinsic origin originates from the Berry phase of electrons in static states, whereas the extrinsic origin needs the scattering effects of electrons.
Thus, we confirmed that the intrinsic and extrinsic origins evaluated by our method exhibit reasonable behavior regarding presence or absence of the electron scattering.
It should be noted that the low value of the side-jump contribution for $S=1$ is ascribed to the artificially small value of the imaginary part in the Green's functions.

With respect to the results for $S=0.99$, we compare the obtained intrinsic and side-jump parts with a previous study assuming the dilute impurity limit; in that study\cite{PhysRevLett.107.106601}, each part of $\sigma_{xy}$ was calculated by substituting the electronic structure of the ordered alloys into the result of a perturbation analysis\cite{PhysRevLett.105.036601}.
The results for the two parts are similar between the two approaches for both alloys.
In particular, our results also indicate that the relative magnitudes of these two parts are different between the FePt and FePd alloys, similar to previous studies\cite{PhysRevLett.107.106601,PhysRevLett.104.076402,PhysRevLett.109.066402}.% in common with the result of previous studies.%; this result is also indicated by experimental studies [].
%It should be noted that, different from the diverged contribution of skew-scattering in the previous studies, our method can evaluate the finite contribution of skew-scattering as well as the other two origins.

\begin{figure}[t]
\includegraphics[width=0.4\textwidth]{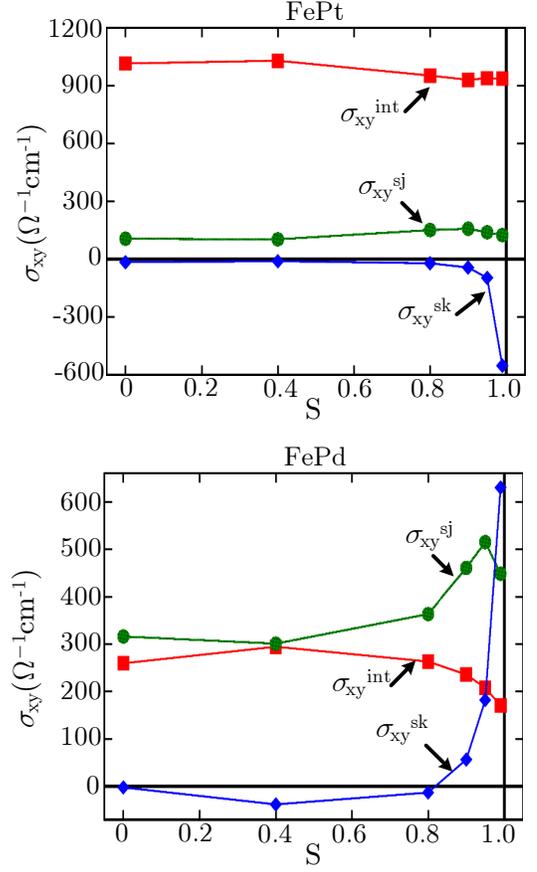}
\caption{$S$ dependence of each part of $\sigma_{xy}$ in FePt and FePd alloys. The parenthetic value of $\sigma_{xy}^{\mathrm{int}}$ denotes the Fermi-sea part expressed in eq. (\ref{eq:16})}
\label{fig:fig4}
\end{figure}

Compared with the previous studies\cite{PhysRevLett.107.106601,PhysRevB.89.014411} based on a perturbation analysis, our calculation method has the advantage of being able to change the strength of electron scattering as a function of $S$. %, which is determined self-consistently by CPA as a function of $S$.
Fig. \ref{fig:fig4} shows the dependence of the estimated values of all parts of $\sigma_{xy}$ on $S (0\le S<1)$ in the FePt and FePd alloys. 
%In this figure, we exclude the result of $S=1$ because
We found that, 1) except for $S\simeq 1$, $\sigma_{xy}$ of FePt is almost dominated by the intrinsic origin, whereas the main origin for FePd is the side-jump contribution, and 2) in the $S\simeq 1$ region, the skew-scattering contribution is predominant for $\sigma_{xy}$ in both alloys. 
When we focus on the $S$ dependence of each origin of $\sigma_{xy}$, the intrinsic and side-jump contributions are almost constant with $S$. 
On the other hand, the skew-scattering part tends to diverge at $S\simeq 1$ and is attenuated as $S$ decreases.

\begin{figure}[t]
\includegraphics[width=0.4\textwidth]{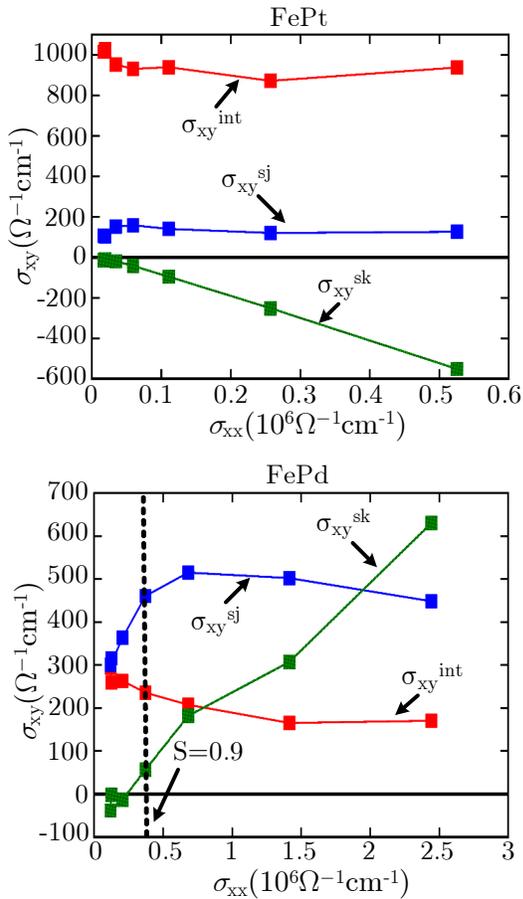}
\caption{Relationship between $\sigma_{xx}$ and each contribution of $\sigma_{xy}$ for different values of $S$. Each superscript (int, sj, and sk) of $\sigma_{xy}$ refers to each part of $\sigma_{xy}$, i.e., intrinsic, side-jump, and skew-scattering.}
\label{fig:fig5}
\end{figure}

%The trigger of these $S$ dependence are well understood from figure \ref{fig:fig5}.
One important result, which indicates the validity of our evaluation method, is the relation between each part of $\sigma_{xy}$ and $\sigma_{xx}$ in FePt and FePd alloys.
Fig. \ref{fig:fig5} shows this relation for the two alloys when $S$ is varied.
A high (low) $\sigma_{xx}$ value is obtained from the alloys having a high (low) $S$.
%The skew-scattering part is nearly proportional to $\sigma_{xx}$, whereas the intrinsic and side-jump contributions tend to be constant as a function of $\sigma_{xx}$ for both alloys.
%These behaviors correspond with those in Eq. (\ref{eq:1}).
The skew-scattering part is nearly proportional to $\sigma_{xx}$ in both alloys,
whereas the intrinsic and side-jump contributions tend to be constant in the high $\sigma_{xx}$ region and vary at low $\sigma_{xx}$.
%On the other hand, the intrinsic and side-jump contributions tend to be constant in the high $\sigma_{xx}$ region and vary in low $\sigma_{xx}$.
In particular, for FePd, $\sigma_{xy}^{\mathrm{int}}$ and $\sigma_{xy}^{\mathrm{sj}}$ show considerable dependencies on $\sigma_{xx}$ in the region of $\sigma_{xx}<0.4\times 10^6 (\Omega^{-1}\mathrm{cm}^{-1})$ $(S<0.9)$.
These behaviors of the intrinsic and side-jump parts are attributed to the change in the band structure due to the disordered alignment of atoms and has little correlation with $\sigma_{xx}$.
These dependencies of each part of $\sigma_{xy}$ on $\sigma_{xx}$ are consistent with Eq. (\ref{eq:00}).
%, which describes the property of each part of $\sigma_{xy}$ for the change of electron scattering.
From this correspondence, we conclude that each calculated contribution of $\sigma_{xy}$ has the adequate properties regarding the strength of electron scattering in these two alloys.
%We confirmed that the estimated each origin obeys
%these different behavior of each origin for $\sigma_{xx}$ correspond with Eq. (\ref{eq:1}).
The behavior of each part of $\sigma_{xy}$ consistent with Eq. (\ref{eq:00}) also explains its dependence on $S$ shown in fig. \ref{fig:fig4} from the fact that $\sigma_{xx}$ has a positive correlation with $S$. %
%only the skew scattering part strongly depends on $S$ in fig. \ref{fig:fig4} because $\sigma_{xx}$ has positive correlation with $S$ in general.
%Consequently,
%In FePd, one can recognize that the intrinsic and side-jump parts vary in low $\sigma_{xx}$ region in fig.\ref{fig:fig5}.
%This behavior is attributed to change of the band structure by $S$ and have little correlation with $\sigma_{xx}$.

\begin{table}[t]
\caption{Contributions of each part of $\sigma_{xy}(\Omega^{-1}\mathrm{cm}^{-1})$ from the conductive process of the vertex correction (VC) and that without the VC when $S = 0.9$ in the FePt and FePd alloys.}
\label{fig:tab2}
\begin{center}
\begin{tabular}[t]{ccc}\hline \hline
FePt & \hspace{0.5em}VC\hspace{0.5em} & w/o VC\\ \hline
$\sigma_{xy}^{\mathrm{int}}$ & 5 & 936 \\
$\sigma_{xy}^{\mathrm{sj}}$ & 32 & 127 \\
$\sigma_{xy}^{\mathrm{sk}}$ & -43 & 0 \\ \hline \hline
\end{tabular}
\vspace{0.5em}
\begin{tabular}[t]{ccc}\hline \hline
FePd & \hspace{0.5em}VC\hspace{0.5em} & w/o VC \\ \hline
$\sigma_{xy}^{\mathrm{int}}$ & 1 & 235 \\
$\sigma_{xy}^{\mathrm{sj}}$ & 27 & 437 \\
$\sigma_{xy}^{\mathrm{sk}}$ & 56 & 0 \\ \hline \hline
\end{tabular}
\end{center}
\end{table}

Finally, we confirmed the contributions of the VC terms for each origin of $\sigma_{xy}$.
Table \ref{fig:tab2} lists each calculated part of $\sigma_{xy}$ from the contribution of the VC terms and without the VC terms in the FePt and FePd alloys. % from the conductive processes with and without the VC when $S = 0.9$ in the FePt and FePd alloys.
The sum of the two contributions corresponds with the total value of each part of $\sigma_{xy}$.
The intrinsic part is mainly dominated by the process without the VC because most parts of its contribution arise without electron scattering.
On the other hand, the VC terms serve a main role in the skew-scattering part owing to the origin of the antisymmetric scattering, which originates from the VC.
The interesting result is that the side-jump contributions consist of both processes, which is the same as the results of previous analyses\cite{PhysRevB.75.045315,PhysRevLett.105.036601}.
From these results, we conclude that the previous separation method for the intrinsic and extrinsic origins of $\sigma_{xy}$ based on the contributions with and without the VC terms, which has been used in existing calculations\cite{PhysRevLett.105.266604,PhysRevB.84.214436,PhysRevB.86.014405,PhysRevB.89.224422,PhysRevB.92.224421}, is justified only in the case where the side-jump contribution is sufficiently small compared with the other two parts.

\section{Summary}
We presented a first-principles technique to evaluate the intrinsic, side-jump, and skew-scattering parts of $\sigma_{xy}$ using the TB-LMTO method.
We performed this separation by distinguishing the intraband and interband elements of the velocity operator $v_{x}$ and $v_{y}$ in the Kubo--Bastin formula.
The application details of the above framework to the TB-LMTO method within the CPA were also presented.

%Using the above framework, the previous calculation procedure of the total $\sigma_{xy}$ in TB-LMTO method was developed into further evaluating each parts of the $\sigma_{xy}$ of disordered alloys in the frame of CPA.
We applied our calculation method to disordered FePt, and FePd alloys, where the relaxation time of electrons was changed as a function of $S$.
We found that, 1) only the intrinsic contribution has considerable value in the ordered phase of the two alloys, and 2) the intrinsic and extrinsic parts respectively exhibit continuous and noncontiguous behavior from the order to disorder transition in both alloys.
These results agreed with their origin, where only the extrinsic origin originates from electron scattering.
When the order parameter $S$ was widely changed, the skew-scattering contribution was proportional to $\sigma_{xx}$, whereas the intrinsic and side-jump parts were almost constant with respect to $\sigma_{xx}$ in both alloys; these results were consistent with Eq. (\ref{eq:00}).
%From these results % of (A) and (B), we confirmed that each part estimated using this method had the adequate physical properties regarding the dependence on electron scattering.
In addition, we discussed the previous separation approach, which distinguishes the intrinsic and extrinsic parts of $\sigma_{xy}$ on the basis of the presence of the VC terms, and we found that the validity of this method was limited to the case where the side-jump contribution was sufficiently small.

Consequently, we demonstrated that our presented evaluation method is applicable to more types of disordered systems than the previous one.
This introduced technique is helpful for understanding the physical origin of $\sigma_{xy}$ more deeply in real alloys.

\begin{acknowledgments}
This work was supported by a Grant-in-Aid from the Japan Society for the Promotion of Science (JSPS) Fellows (No. 25-3505) and KAKENHI from JSPS (No. 16K06702). 
\end{acknowledgments}

\appendix
\section{Specific expression for $\sigma_{xy}^{(1),(2)}$ in our calculation method}
The specific form of $\sigma_{xy}^{(1),(2)}$ that we used is as follows:
\begin{align}
&\sigma_{xy}^{(1),\alpha}\left(\left(v_{\mathrm{eff}}^{\mathrm{D}}(E_{\mathrm{F}})\right)_x^{\alpha},\left(v_{\mathrm{eff}}^{\mathrm{D}}(E_{\mathrm{F}})\right)_y^{\alpha}\right) \notag \\
= & \sigma_0 \mathrm{Tr} \left. \left\{\left(v_{\mathrm{eff}}^{\mathrm{D}}(E_{\mathrm{F}})\right)^{\alpha}_{x} \left(\bar{g}_+^{\alpha}(E_{\mathrm{F}}) - \bar{g}_-^{\alpha}(E_{\mathrm{F}})\right) \left(v_{\mathrm{eff}}^{\mathrm{D}}(E_{\mathrm{F}})\right)^{\alpha}_{y}\bar{g}_-^{\alpha}(E_{\mathrm{F}}) \right.\right. \notag \\
&\hspace{1.5em} \left.\left. - \left(v_{\mathrm{eff}}^{\mathrm{D}}(E_{\mathrm{F}})\right)^{\alpha}_{x} \bar{g}_+^{\alpha}(E_{\mathrm{F}}) \left(v_{\mathrm{eff}}^{\mathrm{D}}(E_{\mathrm{F}})\right)^{\alpha}_{y} \left(\bar{g}_+^{\alpha}(E_{\mathrm{F}}) - \bar{g}_-^{\alpha}(E_{\mathrm{F}}) \right)\right\} \right.\notag \\
&+ \sigma_0 \mathrm{Tr} \left\{ (\alpha-\braket{\gamma}) \left\{ 1+S^{\alpha} (\alpha-\braket{\gamma}) \right\}^{-1}\right. \notag \\
 & \hspace{1.5em} \times \left[ \left(v_{\mathrm{eff}}^{\mathrm{D}}(E_{\mathrm{F}})\right)_x^{\alpha} \left(\bar{g}_+^{\alpha}(E_{\mathrm{F}}) - \bar{g}_-^{\alpha}(E_{\mathrm{F}}) \right) \left(v_{\mathrm{eff}}^{\mathrm{D}}(E_{\mathrm{F}})\right)_y^{\alpha}\right. \notag \\
 & \hspace{2.0em} - \left.\left. \left(v_{\mathrm{eff}}^{\mathrm{D}}(E_{\mathrm{F}})\right)_y^{\alpha} \left(\bar{g}_+^{\alpha}(E_{\mathrm{F}}) - \bar{g}_-^{\alpha}(E_{\mathrm{F}}) \right) \left(v_{\mathrm{eff}}^{\mathrm{D}}(E_{\mathrm{F}})\right)_x^{\alpha} \right] \right\}\notag \\
&+ \sigma_{xy}\left(\left(v_{\mathrm{eff}}^{\mathrm{D}}(E_{\mathrm{F}})\right)_x^{\alpha},\left(v_{\mathrm{eff}}^{\mathrm{D}}(E_{\mathrm{F}})\right)_y^{\alpha}\right)_{\mathrm{ver}} ,\label{eq:A1}\\
= & \sigma_{xy}^{(1),\alpha}\left(\left(v_{\mathrm{eff}}^{\mathrm{D}}(E_{\mathrm{F}})\right)_x^{\alpha},\left(v_{\mathrm{eff}}^{\mathrm{D}}(E_{\mathrm{F}})\right)_y^{\alpha}\right)_{\mathrm{coh}} \notag \\
&+ \sigma_{xy}^{(1),\alpha}\left(\left(v_{\mathrm{eff}}^{\mathrm{D}}(E_{\mathrm{F}})\right)_x^{\alpha},\left(v_{\mathrm{eff}}^{\mathrm{D}}(E_{\mathrm{F}})\right)_y^{\alpha}\right)_{\mathrm{ver}} \label{eq:A1.5}
%& \times \left. \left[v^{(\mathrm{D}),\alpha}_{x} \left[g_{+}^{\alpha} -g_{-}^{\alpha}\right] v^{(\mathrm{D},\alpha)}_{y} \right]\right}
%\\\hspace{1em}\hspace{1em}\hspace{1em}\hspace{1em}\hspace{1em}\hspace{1em}\hspace{1em}\hspace{1em}\hspace{1em}-\left(v_{\mathrm{eff}}^{(\mathrm{D})}\right)_{x}(\alpha-\left\langle \gamma\right\rangle )\leftzfa(1+S^{\alpha}\left(\alpha-\left\langle \gamma\right\rangle \right)\right)^{-1}\left(v_{\mathrm{eff}}^{(\mathrm{D})}\right)_{y}\left[\left(g_{+}^{\alpha}\right)^{'}-\left(g_{-}^{\alpha}\right)^{'}\right]\Bigg\}
\end{align}
\begin{align}
 \sigma_{xy}^{(2),\alpha}
= & - \sigma_0\int_{-\infty}^{E_{\mathrm{F}}}\mathrm{d}E \notag \\
&\times \mathrm{Tr}\left\{v_{x}(g_+^{\alpha}(E))'v_{y}^{\alpha}g_+(E)-v_{x}^{\alpha}g_+^{\alpha}(E)v_{y}^{\alpha}(g_+^{\alpha}(E))'\right.\notag \\
&\hspace{2em}-\left.v_{x}(g_-^{\alpha}(E))'v_{y}^{\alpha}g_-(E)-v_{x}^{\alpha}g_-^{\alpha}(E)v_{y}^{\alpha}(g_-^{\alpha}(E))' \right\}\notag \\
 & -\sigma_0 \int_{-\infty}^{E_{\mathrm{F}}} \mathrm{d} E \mathrm{Tr} \left\{ (\alpha-\braket{\gamma})\left(1+S^{\alpha}(\alpha-\braket{\gamma}\right)^{-1} \right. \notag \\
&  \hspace{4em} \times \left[v_x^{\alpha}\left(\left(g_+^{\alpha}(E)\right)' - \left(g_-^{\alpha}(E)\right)'\right) v_y^{\alpha}\right. \notag \\
& \hspace{4.5em} \left. \left. -v_y^{\alpha}\left(\left(g_+^{\alpha}(E)\right)' - \left(g_-^{\alpha}(E)\right)'\right) v_x^{\alpha} \right] \right\},\label{eq:A2}
%- & \sigma_0 \int_{-\infty}^{E_{\mathrm{F}}}\mathrm{d}E\mathrm{Tr}\left\{(\alpha-\braket{\gamma})\left(1+S^{\alpha}\left(\alpha-\braket{\gamma}\right)\right)^{-1} \left(v^{(\mathrm{D}),\alpha}\right)_{x}\left[\left(g_{+}^{\alpha}\right)^{'}-\left(g_{-}^{\alpha}\right)^{'}\right]\left(v^{(\mathrm{D},\alpha)}\right)_{y}\right,
\end{align}
where $\left(v_{\mathrm{eff}}^{\mathrm{D}}(E)\right)_{\mu}^{\alpha}$ is $\left(v_{\mathrm{eff}}^\mathrm{\mathrm{d,(nd)}}(E)\right)_{\mu}^{\alpha}$ (D is either d or nd), and $\braket{\gamma}$ is the configuration average of $\gamma$, defined as $\braket{\gamma_{\bm{R}}} = \sum_Q c_{\bm{R}}^Q\gamma_{\bm{R}}^Q$.
$\sigma_{xy}^{(1),\alpha}\left(\left(v_{\mathrm{eff}}^{\mathrm{D}}(E_{\mathrm{F}})\right)_x^{\alpha},\left(v_{\mathrm{eff}}^{\mathrm{D}}(E_{\mathrm{F}})\right)_y^{\alpha}\right)_{\mathrm{coh[ver]}}$ denote the coherent part [vertex correction(VC)] of Eq. (\ref{eq:A1}). the VC term mainly originates from the interference effect of the Green's functions in the first term of Eq. (\ref{eq:A1}) and represents antisymmetric scattering, which is necessary for the presence of the skew-scattering contributions. 
The specific form of the VC is obtained by replacing $v_{x}^{\alpha}$ and $v_{y}^{\alpha}$ of  $\sigma_{xy}^{(1),\alpha}\left(v_x^{\alpha},v_y^{\alpha}\right)_{\mathrm{ver}}$, which was shown in the previous studies\cite{PhysRevB.73.144421,PhysRevB.89.064405}, by $\left(v_{\mathrm{eff}}^{\mathrm{D}}(E_{\mathrm{F}})\right)_x^{\alpha}$ and $\left(v_{\mathrm{eff}}^{\mathrm{D}}(E_{\mathrm{F}})\right)_y^{\alpha}$. 
In contrast to Eq. (\ref{eq:A1}), the VC terms of the first term in Eq. (\ref{eq:A2}) completely vanish\cite{PhysRevB.89.064405}.
In addition, the VC contributions in the second terms in Eq. (\ref{eq:15}) and (\ref{eq:16}) are disregarded in our method because these contributions only work the symmetric scattering.
This approximation allows the individual configuration averages of $g$ and $\gamma$.
We evaluate each part of $\sigma_{xy}$ by substituting Eq. (\ref{eq:A1}) and Eq. (\ref{eq:A2}) into Eq. (\ref{eq:27})-(\ref{eq:29}).

\section{representation invariance of each mechanism of $\sigma_{xy}$}
This section shows the invariance of the expression of each mechanism term of $\sigma_{xy}$ given by Eq. (\ref{eq:A1}) and Eq. (\ref{eq:A2}) for the arbitrary representation $\alpha$.  
In the TB-LMTO method, relationship of physical quantities between different representations $\alpha$ and $\beta$ is based on following two equations:
\begin{align}
P^{\alpha}(z) = & \left(1+P^{\beta}(z)(\beta-\alpha)\right)^{-1}P^{\beta}(z),\label{eq:B0}\\
S^{\alpha} = & \left(1+S^{\beta}(\beta-\alpha)\right)^{-1}S^{\beta}. \label{eq:B1}
\end{align}
Applying these above relations into eq. (\ref{eq:12}), eq. (\ref{eq:25}), (\ref{eq:25.1}) and $g^{\alpha}(z) = \left(P^{\alpha}(z)-S^{\alpha}\right)^{-1}$, the transformation properties of physical quantities comprising eq. (\ref{eq:A1}) and eq. (\ref{eq:A2}) is given by   
\begin{align}
{v}_{\mu}^{\alpha} = & K^{-1} v_{\mu}^{\beta} \left(K^{\dagger}\right)^{-1}, \label{eq:B2} \\
%\left(\bm{v}_{\mathrm{eff}}^{\mathrm{D}}(E)\right)_{\mu}^{\beta} = & \left(1+S^{\alpha}(\alpha-\beta)\right)^{-1}v_{\mu}^{\alpha}\left(1+(\alpha-\beta)S^{\alpha}\right)^{-1} \label{eq:B3} \\
\left({v}_{\mathrm{eff}}^{\mathrm{D}}(E)\right)_{\mu}^{\alpha} = & K^{-1} \left(v_{\mathrm{eff}}^{\mathrm{D}}(E)\right)_{\mu}^{\beta} \left(K^{\dagger}\right)^{-1}, \label{eq:B3} \\
g_{\pm}^{\alpha}(E) = & K^{\dagger} g_{\pm}^{\beta}(E) K + K^{\dagger}(\beta-\alpha), \label{eq:B4} \\
%= & \left(1+(\alpha-\beta)S^{\alpha}\right)g^{\alpha}(z) \left(1+S^{\alpha}(\alpha-\beta)\right) \notag \\
\left(g_{\pm}^{\alpha}(E)\right)^{'} = & K^{\dagger} \left(g_{\pm}^{\beta}(E)\right)^{'} K, \label{eq:B5}
\end{align}
where $K=\left(1+S^{\beta}(\beta-\alpha)\right)$. From these relations, the conversion of the coherent part of Eq. (\ref{eq:A1}) and Eq. (\ref{eq:A2}) is expressed as
\begin{align}
&\sigma_{xy}^{(1),\alpha}\left(\left(v_{\mathrm{eff}}^{\mathrm{D}}(E)\right)_x^{\alpha},\left(v_{\mathrm{eff}}^{\mathrm{D}}(E)\right)_y^{\alpha}\right)_{\mathrm{coh}} \notag \\
= & \sigma_0 \mathrm{Tr} \left. \left\{\left(v_{\mathrm{eff}}^{\mathrm{D}}(E_{\mathrm{F}})\right)^{\beta}_{x} \left(\bar{g}_+^{\beta}(E_{\mathrm{F}}) - \bar{g}_-^{\beta}(E_{\mathrm{F}})\right) \left(v_{\mathrm{eff}}^{\mathrm{D}}(E_{\mathrm{F}})\right)^{\beta}_{y}\bar{g}_-^{\alpha}(E_{\mathrm{F}}) \right.\right. \notag \\
&\hspace{0.5em} \left.\left. - \left(v_{\mathrm{eff}}^{\mathrm{D}}(E_{\mathrm{F}})\right)^{\beta}_{x} \bar{g}_+^{\beta}(E_{\mathrm{F}}) \left(v_{\mathrm{eff}}^{\mathrm{D}}(E_{\mathrm{F}})\right)^{\beta}_{y} \left(\bar{g}_+^{\beta}(E_{\mathrm{F}}) - \bar{g}_-^{\beta}(E_{\mathrm{F}}) \right)\right\} \right.\notag \\
& + \sigma_0 \mathrm{Tr} \left\{ T \left[\left(v_{\mathrm{eff}}^{\mathrm{D}}(E_{\mathrm{F}})\right)_x^{\beta} \left(g_+^{\beta}(E_{\mathrm{F}})-g_-^{\beta}(E_{\mathrm{F}})\right) \left(v_{\mathrm{eff}}^{\mathrm{D}}(E_{\mathrm{F}})\right)_y^{\beta} \right.\right. \notag \\
& \hspace{3.5em} \left.\left. -\left(v_{\mathrm{eff}}^{\mathrm{D}}(E_{\mathrm{F}})\right)_y^{\beta} \left(g_+^{\beta}(E_{\mathrm{F}})-g_-^{\beta}(E_{\mathrm{F}})\right) \left(v_{\mathrm{eff}}^{\mathrm{D}}(E_{\mathrm{F}})\right)_x^{\beta} \right]\right\}, \label{eq:B6}
\end{align}
\begin{align}
&\sigma_{xy}^{(2),\alpha} = - \sigma_0\int_{-\infty}^{E_{\mathrm{F}}}\mathrm{d}E \notag \\
&\times \mathrm{Tr}\left\{v_{x}(g_+^{\beta}(E))'v_{y}^{\alpha}g_+(E)-v_{x}^{\beta}g_+^{\beta}(E)v_{y}^{\beta}(g_+^{\beta}(E))'\right.\notag \\
&\hspace{2em}-\left.v_{x}(g_-^{\beta}(E))'v_{y}^{\beta}g_-(E)-v_{x}^{\beta}g_-^{\beta}(E)v_{y}^{\beta}(g_-^{\beta}(E))' \right\}\notag \\
& -\sigma_0 \int_{-\infty}^{E_{\mathrm{F}}} \mathrm{d} E \mathrm{Tr} \left\{ T\left[v_x^{\beta}\left(\left(g_+^{\beta}(E)\right)' - \left(g_-^{\beta}(E)\right)'\right) v_y^{\beta}\right. \right. \notag \\
& \hspace{7em} \left. \left. -v_y^{\beta}\left(\left(g_+^{\beta}(E)\right)' - \left(g_-^{\beta}(E)\right)'\right) v_x^{\beta} \right] \right\},\label{eq:B7}
\end{align}
\begin{align}
 T & =  \left(1+(\beta-\alpha)S^{\beta}\right)^{-1} \notag \\
& \times \left[(\beta-\alpha) + (\alpha-\braket{\gamma}) \right.\notag \\
& \hspace{1.5em} \times \left. \left(1+S^{\alpha}(\alpha-\braket{\gamma})\right)^{-1}\left(1+S^{\beta}(\beta-\alpha)\right)^{-1}\right]. \label{eq:B8}
\end{align}  
By multiple application of Eq. (\ref{eq:B1}), it can be shown that $T=(\beta-\braket{\gamma})\left(1+S^{\beta}(\beta-\braket{\gamma}\right)^{-1}$, which proves
\begin{align}
&\sigma_{xy}^{(1),\alpha}\left(\left(v_{\mathrm{eff}}^{\mathrm{D}}(E)\right)_x^{\alpha},\left(v_{\mathrm{eff}}^{\mathrm{D}}(E)\right)_y^{\alpha}\right)_{\mathrm{coh}} \notag \\
 = & \sigma_{xy}^{(1),\beta}\left(\left(v_{\mathrm{eff}}^{\mathrm{D}}(E)\right)_x^{\beta},\left(v_{\mathrm{eff}}^{\mathrm{D}}(E)\right)_y^{\beta}\right)_{\mathrm{coh}}, \label{eq:B9}\\
&\sigma_{xy}^{(2),\alpha} = \sigma_{xy}^{(2),\beta}.\label{eq:B10}
\end{align}
As for the vertex part of the Fermi-surface term, which is the substituted quantity $\sigma_{xy}^{(1),\alpha}\left(v_x^{\alpha} \rightarrow \left(v_{\mathrm{eff}}^{\mathrm{D}}(E_{\mathrm{F}})\right)_x^{\alpha},v_y^{\alpha} \rightarrow \left(v_{\mathrm{eff}}^{\mathrm{D}}(E_{\mathrm{F}})\right)_x^{\alpha} \right)_{\mathrm{ver}}$ and the representation invariance of the original form of  $\sigma_{xy}^{(1),\alpha}\left(v_x^{\alpha},v_y^{\alpha}\right)_{\mathrm{ver}}$ was shown in the previous studies\cite{PhysRevB.73.144421,PhysRevB.89.064405}, the relation: 
\begin{align}
&\sigma_{xy}^{(1),\alpha}\left(\left(v_{\mathrm{eff}}^{\mathrm{D}}(E_{\mathrm{F}})\right)_x^{\alpha},\left(v_{\mathrm{eff}}^{\mathrm{D}}(E_{\mathrm{F}})\right)_y^{\alpha}\right)_{\mathrm{ver}} \notag \\
= & \sigma_{xy}^{(1),\beta}\left(\left(v_{\mathrm{eff}}^{\mathrm{D}}(E_{\mathrm{F}})\right)_x^{\beta},\left(v_{\mathrm{eff}}^{\mathrm{D}}(E_{\mathrm{F}})\right)_y^{\beta}\right)_{\mathrm{ver}}\label{eq:B10}
\end{align}
can be obtained by the same way as shown in the previous study\cite{PhysRevB.89.064405}, which revealed the representation invariance of $\sigma_{xy}^{(1),\alpha}\left(v_x^{\alpha},v_y^{\alpha}\right)_{\mathrm{ver}}$. 
This treatment for $\sigma_{xy}^{(1),\alpha}\left(\left(v_{\mathrm{eff}}^{\mathrm{D}}(E)\right)_x^{\alpha},\left(v_{\mathrm{eff}}^{\mathrm{D}}(E)\right)_y^{\alpha}\right)_{\mathrm{ver}}$ is allowed by the common transformation properties between $v_{\mu}^{\alpha}$ and  $\left(v_{\mathrm{eff}}^{\mathrm{D}}(E)\right)_{\mu}^{\alpha}$ as shown in Eq. (\ref{eq:B2}) and Eq. (\ref{eq:B3}). 

\bibliography{template}
%\bibliography{D:/reference/library}

\end{document}